\documentclass[12pt]{iopart}
\usepackage{iopams}  
\usepackage{graphicx}
\usepackage[dvipsnames]{xcolor}

\begin{document}
	
	
	\title{The stochastic gravitational wave background from magnetars.}
\author{Sourav Roy Chowdhury}

\address{Research Institute of Physics, Southern Federal University, 344090 Rostov on Don, Russia.}
\ead{roic@sfedu.ru}
	
\author{Maxim Khlopov}
\ead{khlopov@apc.in2p3.fr}
\address{Research Institute of Physics, Southern Federal University, 344090 Rostov on Don, Russia.\\ APC Laboratory 10, rue Alice Domon et L\'eonie Duquet, 75205 Paris Cedex 13, France. \\Center for Cosmoaprticle Physics Cosmion, Moscow State Engineering Physics Institute, National ResearchNuclear University “MEPHI”, 31 Kashirskoe Chaussee, 115409 Moscow, Russia}

	\date{\today}
	
	\begin{abstract} Magnetars have already been a potential candidate as gravitational wave sources that could be detected by current and future terrestrial as well as ground based gravitational wave detectors. In this article we focus on the gravitational wave emission from the distorted rotating neutron stars. The deformation is assumed to be symmetric around an axis that is perpendicular to the rotation axis. The form is applied in the context of a neutron star whose magnetic field has been deformed by its own. By introducing the effects from all magnetars in the Universe, based on various proposed magnetic field configurations, such as poloidal, toroidal; stochastic gravitational wave background (SGWB) can be generated. We choose to figure out exactly how the observations of the stochastic gravitational-wave background should be used to understand much more about physics correlated with the magnetar behavior, based on the restriction on the eccentricity of the magnetar. 
	\end{abstract}

	\maketitle
	
	
	\section{Introduction}	
	The detection of gravitational waves (GWs) was achieved by the Advanced LIGO and VIRGO team \cite{abbott9,abbott8,abbott7,Ligo1,abbott1,abbott2,abbott3,Abadie}, and the presence of GWs is estimated by Einstein’s general relativity. This significant step forward in GW research provides an opportunity to investigate gravity in intensely dynamic and robust regions. A few events, such as black hole--black hole mergers, neutron star--neutron star mergers, or black hole--neutron star mergers \cite{abbott4,abbott5,abbott6}, are considered to have a different class of results. When used in conjunction with some electromagnetic contemporaries, it may be possible to better understand physics, particularly in the most extreme regimes of gravitational fields, densities, magnetic fields, and other variables.

	The contributions of superposition from many independent and unresolved GW sources are expected to produce the stochastic gravitational wave background (SGWB). The SGWB could be cosmological, arising in various inflationary models \cite{Grishchuk,Starobinskii,Barnaby} or cosmic string models \cite{Caldwell,Damour1,Vuk}; it could be astrophysical because of the superposition of waves generated by many astrophysical sources, e.g., compact binary coalescences (CBCs) \cite{Regimbau2,Regimbau3,Rosado2,Vuk1}; or it could be from neutron stars \cite{Regimbau,Rosado3} (quadrupole emission, including magnetars) or initial instabilities \cite{Owen,Howell,Ferrari}. Binary neutron star mergers with multi messengers, such as GWs and electromagnetic counterparts of the signals, are studied to estimate the Hubble constant, which measures the rate of expansion of the Universe,
	and to study the behaviour of the matter which are denser than an atomic nucleus \cite{Dietrich}. The Hubble constant is estimated to be around 70 km/sec/Mpc. This figure is in line with previous estimates of Planck 2015 results \cite{abbott11, Coughlin}. To explain the observed features of soft gamma repeaters (SGRs) and anomalous X-ray pulsars (AXPs), magnetars, neutron stars with a strong magnetic field were first proposed \cite{Duncan}. The strong magnetic field introduced induced a quadrupolar deformation in the magnetar structure, generating GWs during its rapid spinning, in addition to driving the powerful electromagnetic radiation that enabled observation of these objects.

	In the frequency bandwidth of LIGO, VIRGO, and KARGA \cite{abbott8,abbott10, Kuroda1} interferometric detectors, rapidly rotating neutron stars could be a promising source of consecutive gravitational waves. Second- and third-generation GW detectors could be more beneficial to detect the SGWB caused by magnetars. Extensive studies are in the following references \cite{Vuk2,Regimbau4,Vuk3}. In Advanced LIGO's second observing run, the LIGO investigated short- and intermediate-duration GW signals from four magnetar bursts. They found no proof of the signal and ascertained a constraint on the root sum square of the maximum dimensionless strain (from incoming intermediate-duration GWs of a span of the spectrum). Based on optimal orientation, the collaborations place upper bounds on the isotropic GW energy for a magnetar with a known distance (SGR 1806220 (8.7 kpc))\cite{abbott7}. Macquet et al. \cite{Macquet} recommended a new GW search algorithm.

	A stationary rotating body that is perfectly symmetric does not emit GWs. A pulsar accordingly must deviate from axisymmetry to radiate gravitationally. If an axisymmetric, rotating body radiates gravitationally, its axis of rotation cannot coincide with the axis attributed to the symmetry \cite{Zimmermann}. The rotation axis of a neutron star might not even coincide with a principal axis of its moment of inertia due to its violent formation (supernova) or its environment, and the star may precess \cite{Pines}. If the magnetic axis and the rotational axis do not line up, it is assumed that the magnetic pressure can be the origin of the contortion of the star. The magnetic fields of neutron stars are known to be substantial. Another reason for considering the mechanism of asymmetries is the advancement of non-axisymmetric instabilities. The gravitational radiation reaction \cite{Schurtz} or nuclear matter viscosity \cite{Bocquet,Bonazzola} drive rapid rotation in the neutron stars. An internal magnetic field that causes the deformation is also associated with the period derivative ($\dot{P}$) of the pulsar \cite{Gal}. SGWB generated from magnetic deformation of newly born magnetars was studied by Cheng et al. \cite{Cheng} in the context of the growing tilt angle between the star's spin and magnetic axis due to the internal viscosity effects. The magnetars with ultra-strong toroidal magnetic fields contribute to the GWs' background spectra. The background radiation above 1000 Hz is suppressed, unlike the background spectra estimated by assuming constant tilt angles at $ \pi / 2 $. 
	
	Gravitational radiation is produced at two frequencies in the general case: $\Omega$ and 2$\Omega$, where $\Omega$ is the rotational frequency. Instead of considering analytical expressions of GWs, some have provided the amplitudes $h_+$ and $h_{\times}$ directly from the Riemann tensor \cite{Bonazzola}.

	In addition, to accomplish powerful electromagnetic radiation, the observations of such strongly magnetized compact objects are encouraged. SGWB is formed by the superposition of GWs produced by all the magnetars in the Universe. We classified our study in the following manners: in Section \ref{model}, the model of the SGWB has been modified with sophistication. Different magnetic field configurations (poloidal and toroidal field configuration) within the magnetars have been studied based on ellipticities of the objects in Section \ref{mag}, and we conclude and discuss in Section \ref{Con}.
	
	\section{SGWB Model} \label{model}
	
	Stochastic gravitational radiation background is a spontaneous GW signal accumulated by many weak, unresolved gravity wave sources. It resembles the cosmic microwave background radiation (CMBR), which appears to be a stochastic background of electromagnetic radiation \cite{Kolb} in several ways. The spectral features of the gravitational background can be characterized in the same fashion as the CMBR by defining how well the energy is distributed in frequency. The normalized GW energy spectrum is commonly used to describe the SGWB \cite{Allen}, as follows.
	
	\begin{equation}
	\Omega_{gw}(f) = \frac{1}{\rho_c} \frac{d \rho_{gw}}{d \ln f},
	\end{equation}
	where $d \rho_{gw} $ is the energy density in the range of frequency band $f \rightarrow f +df$. $\rho_c$ is the critical energy density (today), related with the Hubble constant ($H_0$) as follows:
	
	\begin{equation}
	\rho_c = \frac{3 H_0^2 c^2}{8 \pi G}.
	\end{equation}
	where G and c are the Newtonian gravitational constants and light speed, respectively.

	Energy $ (e) $ propagating over an area $ (A) $  in time $ (t) $ can be related by energy flux $ (F)  = e/ At$. Now, from the definition of the SGWB, the energy density spectrum becomes as follows:
	
	\begin{equation}
	\Omega_{gw}(f) = \frac{1}{\rho_c} \frac{f}{c} \frac{dF}{df}. \label{3}
	\end{equation} 
	
	Again, the total flux can be written as the integral of fluxes emanating from all sources:
	
	\begin{equation}
	\frac{dF}{df} = \int_{0}^{\infty} \frac{R(z) dz}{ 4 \pi d_L^2(z)} \frac{1}{(1+z)} \frac{d E_{gw}}{df}\bigg|_{f(1+z)}. \label{4}
	\end{equation}   
	
	Here, $ R(z) $ is the rate of magnetars as a function of redshift $ (z) $. The gravitational wave energy spectrum is  $ \frac{d E_{gw}}{df} $ and the luminosity distance $ d_L(z) $ is correlated with the proper distance $ r(z) $ as $d_L(z)=(1+z)r(z)  $. 
	
	The $ (1+z) $ factor accounts for cosmic advancement and transforms the time in the source picture to the detector picture. Eventually, the redshift influence of the comoving volume is captured by $ E(\Omega_m, \Omega_\lambda, z) $ as follows:
	
	\begin{equation}
	E(\Omega_m, \Omega_\lambda, z) = \sqrt{\Omega_m (1+z)^3+\Omega_\lambda}, 
	\end{equation}
	where $ \Omega_m $ and $ \Omega_\lambda $ are the energy density in matter and in dark energy. $R(z)$ is the rate of magnetars or the GW sources in terms of red shift $ (z) $ as observed on Earth, measured in local time. It is correlated with the magnetars' rate per unit comoving volume $ R_V(z) $, as follows:
	
	\begin{equation}
	R(z) = \frac{4 \pi d_L^2 c}{H_0} \frac{R_V(z)}{E(\Omega_m, \Omega_\lambda, z)}, \label{6}
	\end{equation}
	where $ R_V(z) $ is measured in the source frame time. Triaxial-shaped rotating neutron stars (NSs) have a time-varying quadrupole moment, which causes them to emit GWs at the double rate of rotational frequency. The complete spectral gravitational energy emitted by an NS \cite{Vuk3} with a rotational period of $ P_0 $ that slows down due to magnetic dipole torques and GW emission is given by 
	
	\begin{equation}
	\frac{d E_{gw}}{df}  = I f^3 \pi ^2 \bigg(\frac{5c^2 R^6 B^2}{192 \pi ^2 G I^2 \epsilon ^2} + f^2\bigg)^{-1},
	\end{equation}  
	where $ R $ is the magnetar radius, $ B $ is the magnetic field, $ I $ its moment of inertia, and $ \epsilon $ is the ellipticity of the magnetar. The first term in the bracket comes from the electromagnetic radiation's angular acceleration, while the second term comes from the contribution of GW emission acceleration.

	With the help of Eq. (\ref{4}) and Eq. (\ref{6}), we therefore rewrite Eq. (\ref{3}) as follows:
	
	\begin{equation}
	\Omega_{gw}(f) = \frac{f}{\rho_c H_0} \int_{0}^{\infty}  \frac{R_V(z) dz}{(1+z)E(\Omega_m, \Omega_\lambda, z)} \frac{d E_{gw}}{df}\bigg|_{f(1+z)}.
	\end{equation}

	\section{Configuration of Magnetars} \label{mag}
	
	The magnetic field and the ellipticity of the magnetars are related separately, depending on their field configurations. SGWB searches can be used to examine these various field structures and the physics that also prompt them---in particular, a poloidal magnetic field expanded from the interior to the exterior of the magnetar. A toroidal magnetic field aspect has also been suggested to occur only within the torus-shaped region of the magnetar.

	\subsection{Poloidal Field Configuration}
	
	The variables of self-consistent models of magnetized neutron stars can be used to estimate the value of B for distinct models of magnetic field distributions. The magnetic field distribution has a significant impression on the disturbance at a specified magnetic dipole moment. In comparison to the uniformly magnetized ideal conductor case, a stochastic magnetic field or a stellar superconductor interior varies significantly with different parameters, which could lead to GWs. 
	
	Bonazzola and Gourgoulhon \cite{Bonazzola} proposed that the ellipticity of the poloidal field configuration is 
	\begin{equation}
	\epsilon = \beta \left(\frac{R^8  B^2}{4 G I^2}\right).
	\end{equation}
	Here, $ \beta $ is a dimensionless quantity. This quantity is taken into consideration in the equation of state and magnetic field geometry. $ R $ is the radius of the magnetar, $ B $ is the magnetic field associated with the corresponding magnetar, and the moment of inertia of the body is $ I $.
	\begin{figure}[h!]
		\centering
		\includegraphics[scale=.5]{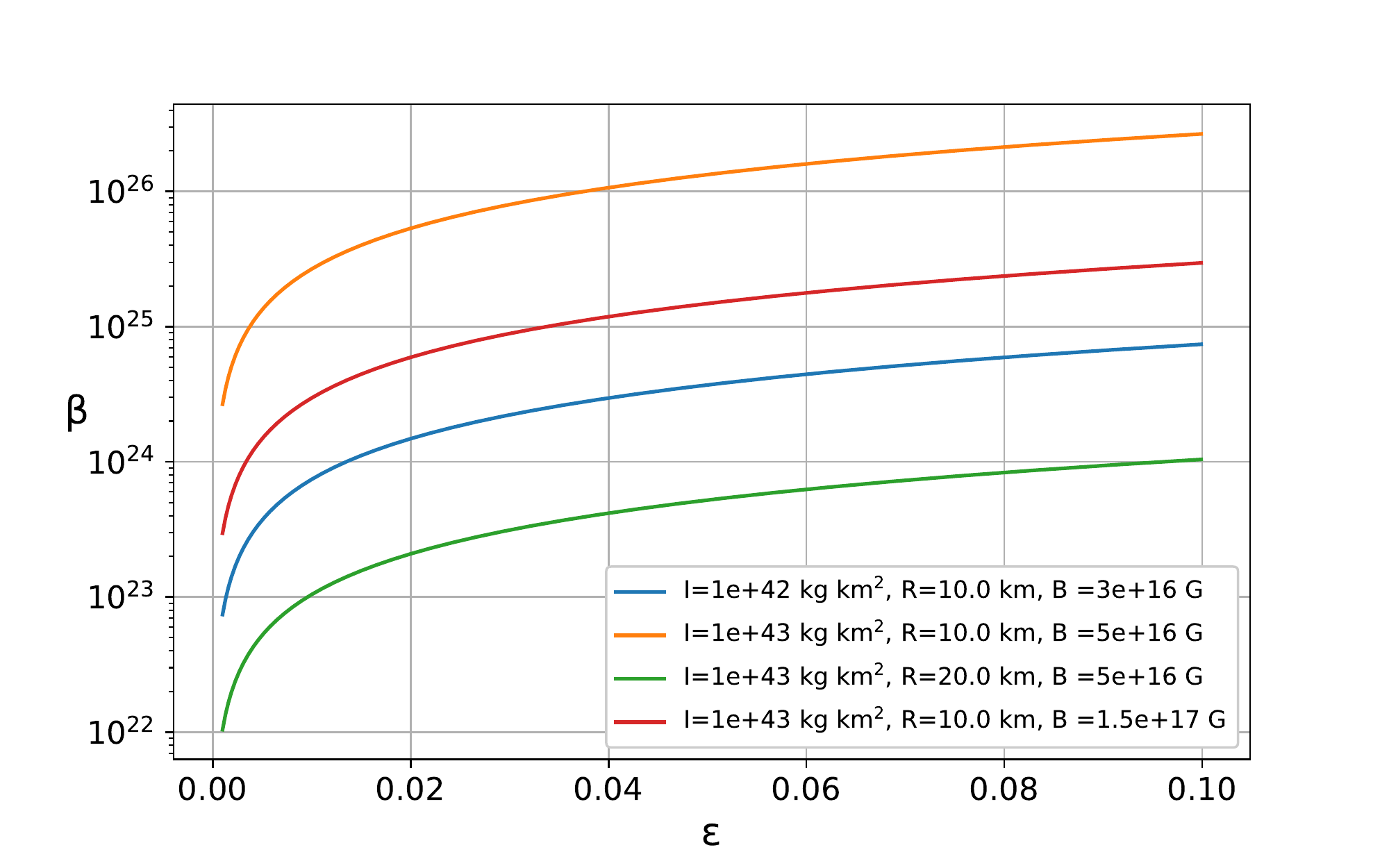}
		\caption{ 
			Expected evolution of the dimensionless parameter $(\beta)$ with ellipticity $(\epsilon)$ for the free parameters radius $ (R) $, magnetic field $ (B) $, and moment of inertia $ (I) $. The parametric values
			are exhibited in the figure.}\label{fig1}
	\end{figure}	
	
	A slightly higher ellipticity leads to an adequately high distortion parameter, and the values are almost stagnant for a range of ellipticity, although they increase rapidly for the lower range.

	\begin{figure}[h!]
		\centering
		\includegraphics[scale=.5]{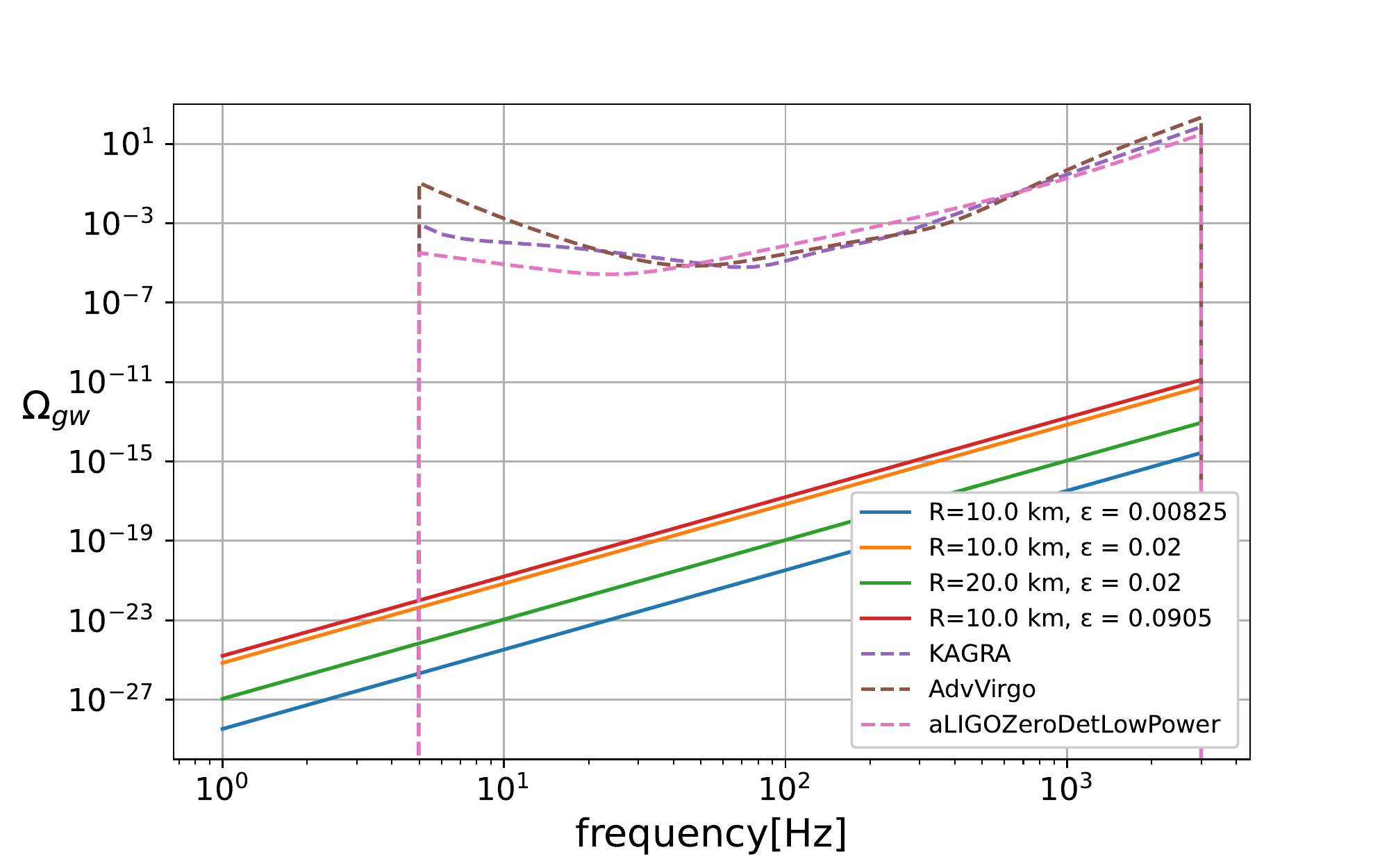} 
		\caption{ $ \Omega_{gw}(f) $ for different parametric preferences for the magnetar, produced SGWB high frequency model of the poloidal field configuration. The free parameters are $\epsilon$ and $ R $, respectively. Furthermore, $ R $ is in km, as shown in the figure. }\label{fig2}
	\end{figure}	
	
	The energy spectrum corresponding to the configuration thus depends on the distortion parameters and the associated magnetic field.

	\subsection{ Twisted-Torus Magnetic Field Configuration}
	
	Braithwaite and Spruit \cite{Braith} developed the twisted-torus magnetic field configuration. It is claimed that instead of the pure poloidal field configuration, the twisted-torus magnetic field configuration has a universal equilibrium structure of the magnetic field, which is potentially stable. A toroidal magnetic field is closed in the interior of the magnetars in the twisted-torus field structure. However, the deformation is mainly dependent on the magnetic field. Huge deformations can be expected if the magnetic field is intense enough during the early stellar stages.

	The poloidal field becomes stabilized and twisted by the toroidal field. The ellipticity can be modeled with the help of a dimensionless parameter, which depends only on the equation of state and field geometry, as follows:
	\begin{equation}
	\epsilon = k \left(\frac{B}{10^{15}}\right)^2 \times 10^{-6}.
	\end{equation}
	\begin{figure}[htp!]
		\centering
		\includegraphics[scale=.5]{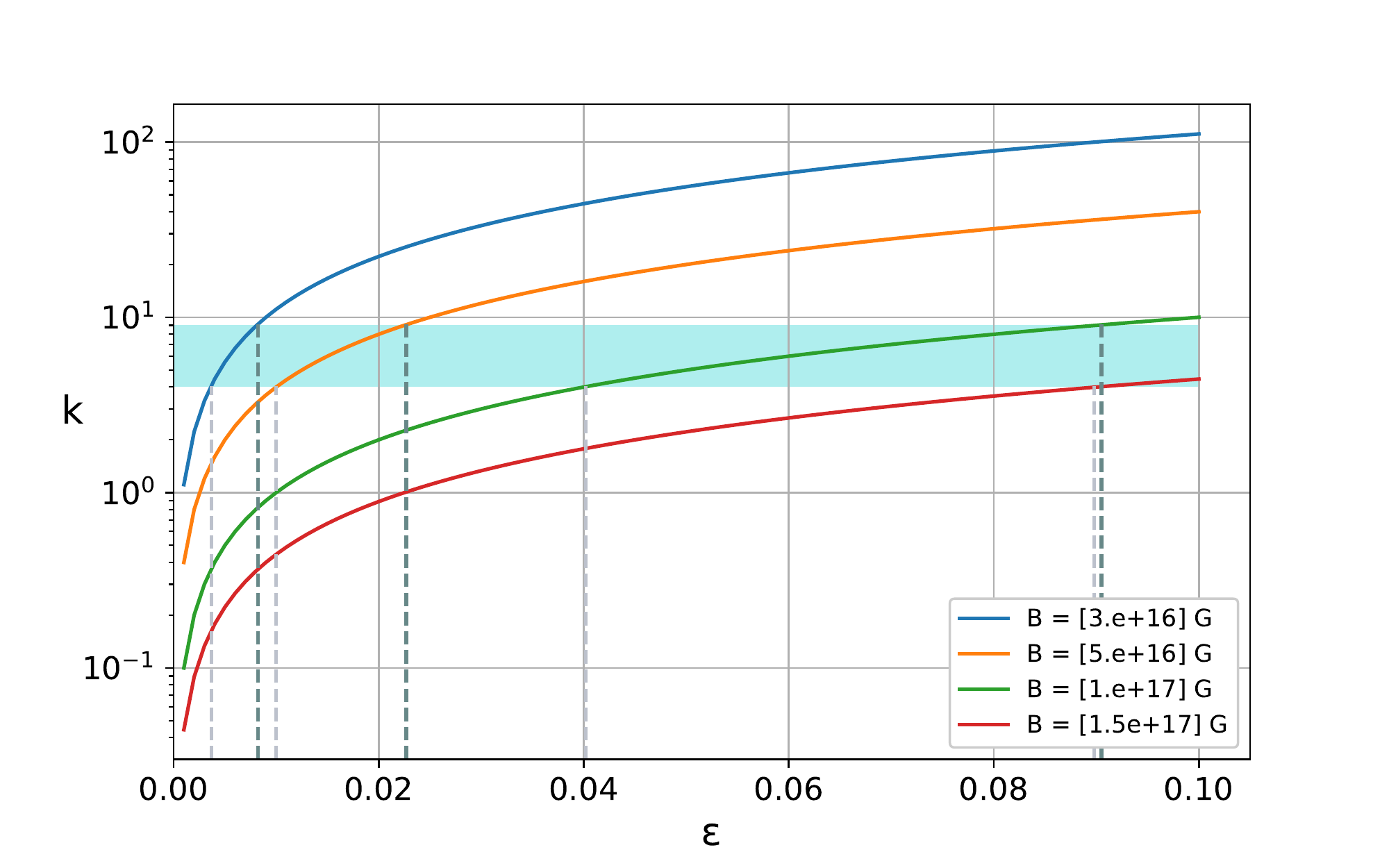}
		\caption{ The dimensionless parameter $(k)$ versus the ellipticity $(\epsilon)$ for the magnetars with different strengths. The only free parameter here is magnetic field $ (B) $, same as in Figure \ref{fig1}. The parametric values are shown in the figure. The range of $ \epsilon $ for being a realistic stellar model for each strength is marked by dashed lines. }\label{fig3}
	\end{figure}	
	
	According to Ciolfi et al. \cite{Ciolfi} for the twisted-torus field configuration, for compact stars in the realistic regime, the dimensionless parameter $ (k) $ must lie within the range $ 4 \rightarrow 9 $. However, from the Akmal--Pandharipande--Ravenhall equation of state, it is clear $ k = 4 $ provides small compactness \cite{Akmal}, while Glendnning \cite{Glendnning} claimed that the significant compactness could arise around $ k = 9 $.

	The emission of GWs from a distorted star is mainly based on its quadrupole ellipticity. Highly deformed, eccentric stellar bodies offer a lower gravitational wave energy spectrum. Even with a higher radius, they can reduce rapidly.
	
	\begin{figure}[htp!]
		\centering
		\includegraphics[scale=.5]{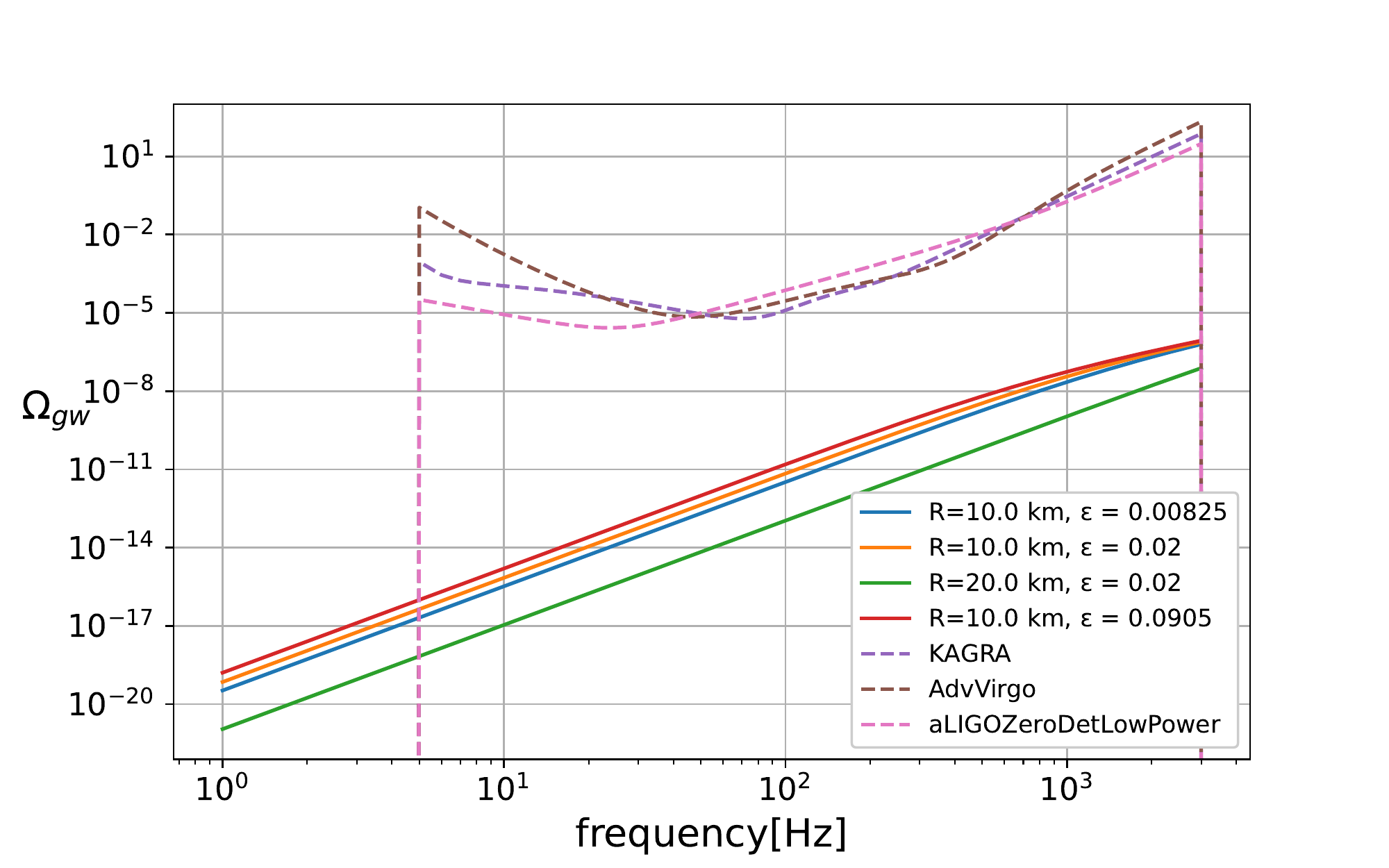}
		\caption{ Variation of $ \Omega_{gw}(f) $ for different parametric preferences for the magnetar producing SGWB from the high-frequency model of the twisted-torus magnetic field configuration. Here, $\epsilon$ and $ R $ are the free parameters, respectively, and $ R $ are measured in $ km $. Values of $ \epsilon $ are chosen within the realist stellar range referred in Figure \ref{fig3}.}\label{fig4}
	\end{figure}

	\section{Conclusions} \label{Con}
	
	In this article, the magnetars are categorized based on their ellipticity. We focused on the finite values (not in a shallow regime) of eccentricities and the corresponding parameters in the range of realistic stellar models.
	
	The strong magnetic field anticipates the deformation of the quadrupolar distortion in the magnetar structure, developing GWs during its rapid spinning. The sign of the quadrupole ellipticity is essential.  $\epsilon < 0$ (corresponds to a prolate structure)  could be the spin-flip process. It can introduce viscous forces. The angle between the magnetic axis and the rotation axis developed on a dissipation time-scale becomes orthogonal. This procedure would be directly connected to the GW emission \cite{Ciolfi}. The poloidal field appears to produce the oblate stellar configuration, whereas the twisted-torus field seems to make prolate. The ellipticity has always been positive for the poloidal field and dominates over the toroidal field. As a result, the spin-flip formalism is incompatible with twisted-torus structures.

	If the interior of the magnetars is considered to be a type I superconductor, emitting a magnetic field from a part of the star, the distortion parameters are much higher. In such cases, more significant distortions are expected, with an overall low magnetic dipole moment. Extremely high stresses in the star's crust could also contribute to higher $ \beta $ values in the type-II superconducting interior. In the context of a poloidally dominated magnetic field structure, it is thus essential to incorporate SGWB constraints on $ B $ and thus obtain information about the equation of state of the magnetars. 
	
	Variations of the dimensionless parameter ($ \beta $  and $ k $, respectively) of the poloidal field configuration and twisted-torus field configuration with the ellipticity are shown in Figure \ref{fig1} and Figure \ref{fig3}, respectively. In both configurations, the values of $ \beta $  and $ k $ are effectively very low for very small values of ellipticity ($ < 10^{-3} $) and become almost saturated with higher values of $ \epsilon $. The blue band in Figure \ref{fig3} defines the realistic regime of $ k $ for the compact stars. Based on Ciolfi et al.'s \cite{Ciolfi} claim that $ 4 \leq k \leq 9 $, the variation of ellipticity is shown for different values of magnetic field $ (B) $; Figure \ref{fig4}. For $ B = 1.0 \times 10^{17}~G $, the $ \epsilon $ of the realistic compact stars has a range $ 0.0402 \rightarrow 0.0905 $, whereas the range is $ 0.0102 \rightarrow 0.0227 $ for $ B = 0.5 \times 10^{17}~G $, and  $ 0.0037 \leq \epsilon \leq 0.0082 $ for $ B = 0.3 \times 10^{17}~G $. From the above analysis, it is clear that higher deformation requires a hierarchically higher magnetic field.

	Variation of $ \Omega_{gw}(f) $ based on the free parameters and dimensionless parameter ($ \beta $ and $ k $, respectively) for the poloidal field configuration and twisted-torus field configuration are shown in Figure \ref{fig2} and Figure \ref{fig4}, respectively. It is found that the normalized energy spectrum for the twisted-torus configuration is higher compared to the poloidal configuration. It makes the possibility of getting twisted-torus configured magnetars higher than the poloidal configuration. It is also interesting to note that the $ \Omega_{gw} $ decreases sufficiently for a higher radius.

	\begin{figure}[h!]
		\centering
		\includegraphics[scale=.5]{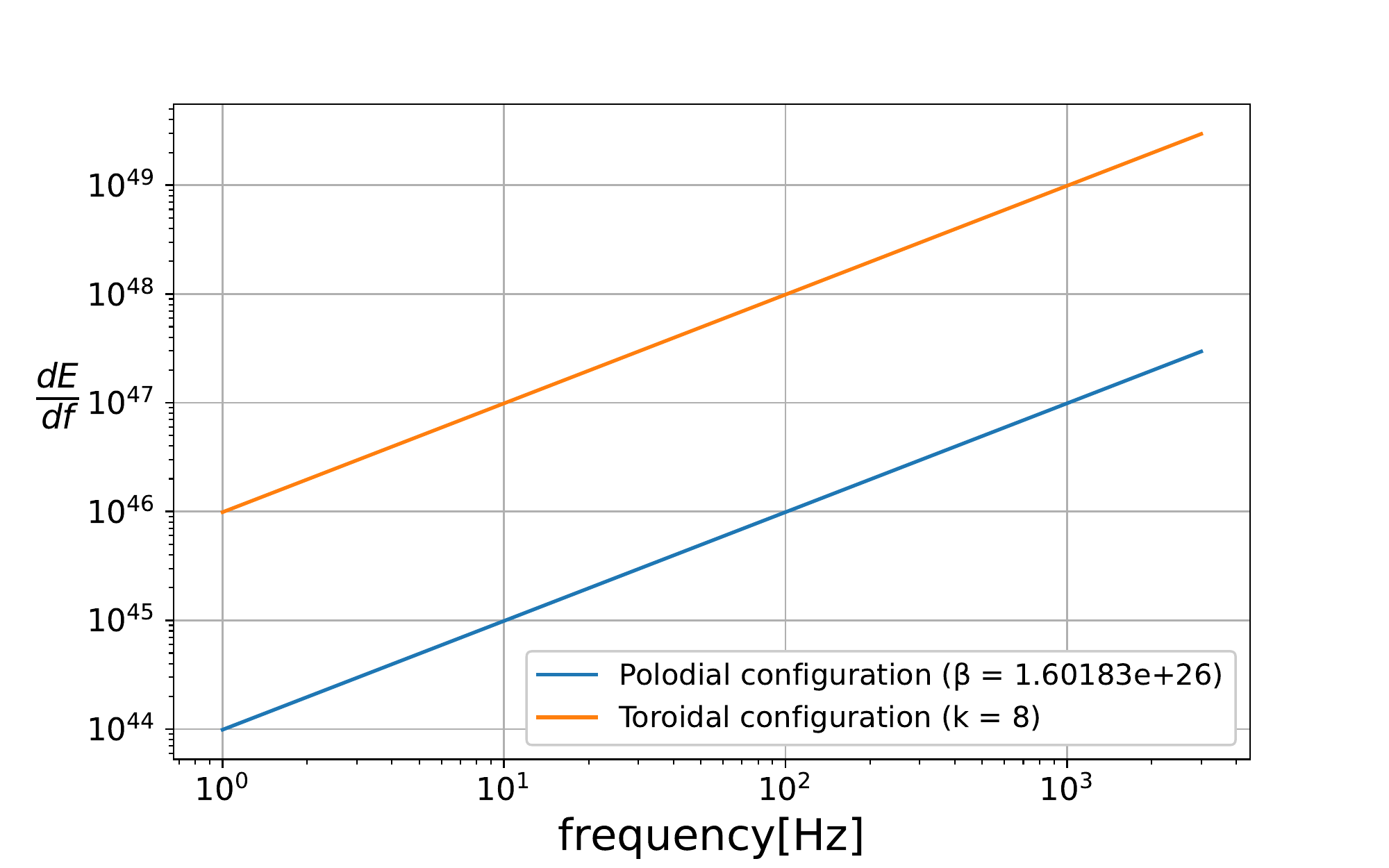}
		\caption{ The expected evolution of the GW energy spectra emitted by the magnetars for different field configuration with frequency for a specific parameter's space. Overall, the energy spectrum tends to increase rapidly as frequency increases. The orange and blue lines represent the results calculated by involving parameters $ I $ = 10$ ^{43}$ kg km$^2$, $ R $ = 10 km, $ \epsilon $ = 0.02.}\label{fig5}
	\end{figure}	
	
	Figure \ref{fig5} also recommends that the energy spectrum for twisted-torus field configuration be higher than the poloidal field configuration. It registers that for the same structural parameter, twisted-torus field configured magnetars can easily be found.
	
	The predicted energy spectrum is less than the so far detected energy spectrum. Nevertheless, for both the configurations, higher ellipticity makes the system more plausible to find. Recently, the collaborations have been searching for the quasi-monochromatic gravitational wave signals from the energetic young X-ray pulsar PSR J0537-6910 \cite{abbott9}. The upcoming presentations of the Advanced LIGO, VIRGO, or KAGRA are awaited to detect the GWs from magnetars. This study can finally affirm that the SGWB energy spectrum for the magnetars with toroidal configuration will emit more detectable GWs than the poloidal configuration.
 

 
 
 \section{Acknowledgments}The work of S.R.C was supported by the Southern Federal University (SFedU) (grant no. P-VnGr/21-05-IF). S.R.C is also thankful to Ranjini Mondol of IISc, Bangalore for the fruitful discussion to improve the manuscript. The research by M.K.was financially supported by Southern Federal University, 2020 Project VnGr/2020-03-IF. We are also grateful to the referees for their valuable comments which made the manuscript even better.



\section*{References}




\end{document}